\documentclass[prl,twocolumn,showpacs,preprintnumbers,amsmath,amssymb]{revtex4}

\usepackage{graphicx}
\usepackage{dcolumn}
\usepackage{bm}

\def\k{{\bf k}}
\def\r{{\bf r}}
 
\def\cx{\cos k_x}
\def\cy{\cos k_y}
\def\E{{\cal E}}

\begin{document}


\title{Absence of single particle Bose-Einstein condensation at low densities for bosons with correlated-hopping}

\author{Rachel Bendjama}
\author{Brijesh Kumar}
 \email{brijesh.kumar@epfl.ch}
\author{Fr\'ed\'eric Mila}
 \email{frederic.mila@epfl.ch}
\affiliation{Institute of Theoretical Physics, \'Ecole Polytechnique F\'ed\'erale de 
Lausanne, CH 1015 Lausanne, Switzerland.}

\date{\today}

\begin{abstract}
Motivated by the physics of mobile triplets in frustrated quantum magnets, the properties of a two dimensional model of bosons with 
correlated-hopping are investigated. A mean-field analysis reveals the presence of a {\em pairing} phase without single particle Bose-Einstein condensation 
(BEC) at low densities for sufficiently strong correlated-hopping, and of an Ising quantum phase transition towards a BEC phase at larger density. 
The physical arguments supporting the mean-field results and their implications for bosonic and quantum spin systems are discussed.
\end{abstract}

\maketitle
The models of interacting bosons (with or without disorder) have been a subject of 
active research. They are studied for a variety of reasons, coming from different 
experimental 
systems, such as Josephson junction arrays~\cite{JJA}, $^4$He in porous 
media~\cite{He}, disordered films with superconducting and insulating 
phases~\cite{Film}, or more recently in the 
context of atoms trapped on an optical lattice~\cite{Optical}. The interplay 
of interaction, disorder and kinetic energy leads to the ground states that can be 
a superfluid, a Bose glass, a 
Mott insulator or a supersolid~\cite{Fisher,Giamarchi,Sorensen,Freericks,TVR,Troyer,Otterlo}. In the context of spin models too, the 
Schwinger boson mean-field theories provide a useful description of 
magnetism in the bosonic language~\cite{AA,HRK,Chandra,Mila}.

Over the last decade, bosons have also been used in the context of quantum 
magnetism to 
describe the magnetization process of gapped systems with a singlet
ground state such as spin ladders, the triplets induced by the magnetic
being treated as hard-core bosons. These bosons may condense,
leading to the ordering of the transverse component of the spins, but they
might as well undergo a superfluid-insulator transition, leading to
magnetization plateaux~\cite{Rice}. For pure SU(2) interactions, and without 
disorder, the common belief is that the only alternative, not realized so far in quantum
magnets, is a supersolid, i.e. a coexistence of these phases.

In this paper, we propose that there is another possibility, namely a {\em pairing} 
phase without single particle Bose condensation. Our starting point is the 
observation that the effective bosonic model of a frustrated quantum magnet
such as SrCu$_2$(BO$_3$)$_2$~\cite{Kageyama} contains, in addition to the usual kinetic
and potential terms, a correlated-hopping term where a boson can hop only
if there is another boson nearby, and that this term can be the dominant source
of kinetic energy in geometries such as the orthogonal dimer model realized
in SrCu$_2$(BO$_3$)$_2$~\cite{Momoi,review_Miyahara}. While the possibility of bound state formation was
already pointed out in that context, the consequences of the presence of such a term
on the phase diagram at finite densities have not been worked out yet.

For clarity, we concentrate in this paper on a minimal version of the model, but 
we have checked that the conclusions apply to the more realistic model 
derived for SrCu$_2$(BO$_3$)$_2$~\cite{Bendjama}. This model is defined 
on a square lattice by the Hamiltonian
\begin{eqnarray}
H &=& -t \sum_\r\sum_{\delta=\pm x,\pm y} b^\dag_{\r+\delta}b_{\r} 
-\mu \sum_\r n_\r \nonumber \\
&&-t^\prime\sum_\r\sum_{\delta=\pm x}\sum_{\delta^\prime=\pm y} 
n_\r\left\{b^\dag_{\r+\delta}b_{\r+\delta^\prime} + h.c. \right\} 
\label{eq:model}
\end{eqnarray}
where $b^\dag_\r, b_\r$ are boson operators and $n_\r=b^\dag_\r b_\r$.
$t$ and $t'$ are the measures of single particle and correlated-hopping respectively.
A hard core constraint that excludes multiple occupancy should in principle be
included. However, we will concentrate on the low density limit, where this 
constraint is expected to be irrelevant. So in the following we work with regular
(soft core) bosons. 

\begin{figure}
\includegraphics{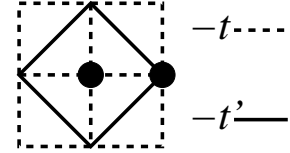}
\label{fig:corr_hop}
\caption{The diagonal bonds ($-t^\prime$) denote a correlated-hopping process, and the nearest neighboring bonds denote the single particle hopping ($-t$).}
\end{figure}
Since the correlated-hopping term in Eq.~(\ref{eq:model}) is quartic in the 
single particle boson operators, the simplest thing to do is a mean-field theory. Since 
the system gains energy through correlated-hopping by having two particles nearby, a 
natural choice for a mean-field is the {\em pairing} amplitude. The particle density and the 
{\em kinetic} amplitudes are the other choices for the mean-fields. In the following, we 
formulate a mean-field theory in terms of these order parameters defined by:
\begin{eqnarray}
\Delta &=& \langle b^\dag_\r b^\dag_{\r\pm\delta} \rangle
\ \ \ (\textrm{pairing amplitude})
\nonumber \\
\kappa &=& \langle b^\dag_\r b_{\r\pm\delta}\rangle,\ 
\kappa^\prime = \langle b^\dag_{\r\pm\delta} b_{\r\pm\delta^\prime}\rangle
\ \ \ (\textrm{kinetic amplitudes}) \nonumber \\
n &=& \langle b^\dag_\r b_\r\rangle \nonumber
\ \ \ (\textrm{particle density})
\end{eqnarray}
where $\delta\neq\delta^\prime$, and $\delta, \delta^\prime=x,y$. The particle density 
is taken to be uniform, and the kinetic amplitudes real. In principle we can allow for an 
internal phase in the pairing amplitude (a non-zero phase between $x$ and $y$ direction bonds)
 like in the mean-field theory of t-J model in the context of the high-T$_c$ cuprates.
 Here, we take the internal phase to be zero (the extended s-wave pairing). The corresponding mean-field Hamiltonian has the following form:
\begin{equation}
H_{MF} = \E_0 + \sum_\k\left\{\xi_\k b^\dag_\k b_\k - \Delta_\k
\left[b^\dag_\k b^\dag_{-\k} + h.c.\right]\right\} \label{eq:model1_Hmf}
\end{equation}
where $\E_0$, $\xi_\k$ and $\Delta_\k$ are given by
\begin{eqnarray}
\E_0 &=& 8t^\prime L [\Delta^2 +\kappa^2+n\kappa^\prime] \nonumber \\
\xi_\k &=& -2(t+4t^\prime\kappa)(\cx+\cy) \nonumber\\ && -
8t^\prime n\cx\cy -(\mu+8t^\prime\kappa^\prime) \nonumber\\
\Delta_\k &=& 4t^\prime\Delta(\cx+\cy) \nonumber
\end{eqnarray}

The Hamiltonian $H_{MF}$ can easily be diagonalized using Bogoliubov transformation 
for bosons. The canonical free energy density for $H_{MF}$ is given as:
\begin{eqnarray}
f &=& \lambda\left(n+\frac{1}{2}\right) + 8t^\prime\left(\Delta^2+\kappa^2\right) 
\nonumber\\ 
&& + \frac{1}{2L}\sum_\k E_\k +\frac{1}{\beta L}\sum_\k\log
\left(1-{\rm e}^{-\beta E_\k}\right) \label{eq:model1_f}
\end{eqnarray}
where $E_\k=\sqrt{\xi_\k^2 - 4\Delta_\k^2}$ is the quasi-particle dispersion, 
$\beta = 1/k_BT$, and $\lambda = \mu +8t^\prime\kappa^\prime$ is the effective chemical 
potential. Re-defining the chemical potential in this way makes $\kappa^\prime$ 
a redundant order parameter in the mean-field theory. 
Note that $\mu$ and $\kappa^\prime$ appear with right 
combination to give $\lambda$ as the new chemical potential in $\xi_\k$. 
Hence, for a given $n$, $f$ is purely a function of $\lambda$, $\kappa$ and $\Delta$.

The self-consistent equations for the order parameters can be written
\begin{eqnarray}
n &=& n_c - \frac{1}{2} + \frac{1}{2L}\sum_\k\frac{\xi_\k}{E_\k}\coth{\frac{\beta E_\k}{2}} 
\label{eq:SCEn_T}\\
\kappa &=& n_c + \frac{1}{4L}\sum_\k\frac{\xi_\k (\cx+\cy)}{E_\k}
\coth{\frac{\beta E_\k}{2}} \label{eq:SCEkappa_T}\\
\Delta &=& n_c + \frac{2t^\prime\Delta}{L}\sum_\k\frac{(\cx+\cy)^2}{E_\k}
\coth{\frac{\beta E_\k}{2}} \label{eq:SCEDelta_T}
\end{eqnarray}
where $n_c$ is the condensate density (the occupancy of the zero energy mode
if any). Since the model is two-dimensional, $n_c=0$ for $T>0$. At $T=0$, $n_c$ may
or may not be zero, and solutions must be searched with two strategies: 
assume $n_c=0$ and solve these equations for the unknowns ($\kappa$, $\Delta$, $\lambda$),
or assume there is a zero energy mode (which fixes $\lambda$), and solve for
the unknowns ($\kappa$, $\Delta$, $n_c$). In that case, the wave-vector corresponding
to the zero-energy mode ($\k = 0$ here) must be excluded from the sum. If several solutions
are found for a given density, the one with lowest energy should be chosen. In practice,
we only found one solution for a given density. In general, these equations are solved by simple iteration. Note however that, when $n_c=0$, 
Eqs.~(\ref{eq:SCEn_T}) and~(\ref{eq:SCEkappa_T}) can be still solved by iteration, but Eq.~(\ref{eq:SCEDelta_T}) needs to be solved for $\Delta$ by some numerical 
method at each step of the iteration. 

\begin{figure}
\includegraphics[width=7.5cm]{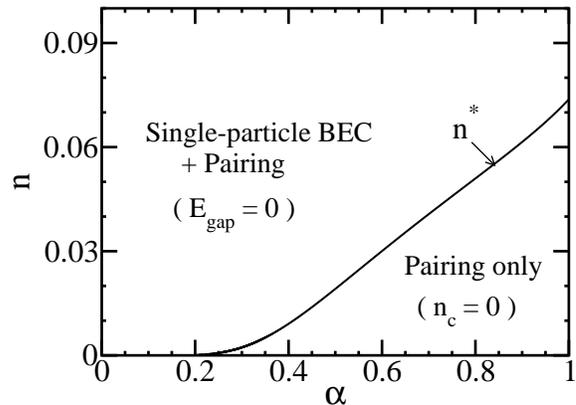} 
\caption{The mean-field quantum phase diagram. $n^*$ is the critical density at which, for a given $\alpha=t^\prime/(t+t^\prime)$, the single particle condensate density $n_c$ becomes zero. 
This marks the onset of the pairing phase with gapped quasi-particle excitations.}
\label{fig:phasediagram}
\end{figure}

Let us first discuss the $T=0$ results. The most remarkable feature is that it turned out
to be impossible to find a solution with a non-zero condensate at low enough density
unless $t'$ is very small. In other words, as soon as the correlated-hopping is not too small, 
there is no single-particle BEC at low density. The critical density $n^*$ below which this is the case is plotted
as a function of $\alpha=t^\prime/(t+t^\prime)$ in Fig.~\ref{fig:phasediagram}.
This figure calls for some comments. First of all, $t'/t$ need not to be large for the 
effect to be observable, which ensures the relevance of the present discussion
for the quantum magnets such as SrCu$_2$(BO$_3$)$_2$. Besides, the critical value  $n^*$ is quite small even when
$t'$ dominates, and our approximation to treat triplets as soft-core bosons is expected
to be good in the whole range of Fig.~\ref{fig:phasediagram}. Finally, densities of
a few percent are definitely accessible in the context of quantum magnets, the density
being equivalent to the magnetization relative to the saturation value.

\begin{figure}
\includegraphics[width=8cm]{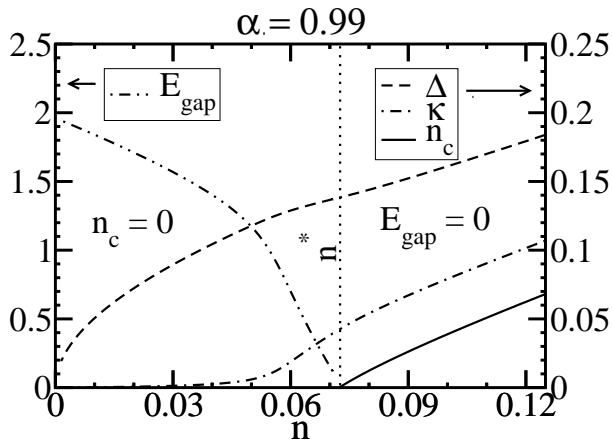}
\caption{The variation of the quasi-particle gap and the mean-field order parameters as a function of $n$ for a fixed $\alpha$. While $n_c\neq 0$ only in the single particle BEC phase 
($n\gtrsim 0.072$), the pairing amplitude $\Delta\neq 0$ in both phases.}
\label{fig:Gap_OP_n}
\end{figure}

\begin{figure}
\includegraphics[width=7.5cm]{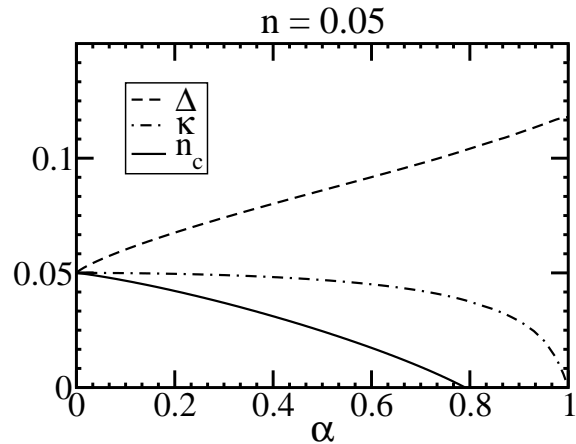}
\caption{The behavior of different order parameters as a function of $\alpha$ for $n=0.05$. 
For $\alpha\gtrsim 0.78$, the condensate density ($n_c$) vanishes, and the system goes into 
         the pairing ground state.}
\label{fig:OP_alpha}
\end{figure}

Next we turn to the nature of this non-BEC phase. Clearly this cannot be a commensurate
insulating phase of the type observed before since it occurs for a range of densities. In fact,
its nature is best revealed by looking at the order parameters. While $n_c=0$ in the pairing phase and E$_{gap}=0$ in the 
single particle BEC phase, the mean-field solution for $\kappa$ and $\Delta$ is non-zero on both sides. In Fig.~\ref{fig:Gap_OP_n}  the behavior of 
$n_c$, $\kappa$ and $\Delta$ is shown as a function of $n$ for $\alpha=0.99$.
It is not surprising that $\Delta$ is non-zero in the single particle BEC phase. In fact, 
the single particle BEC state means $\langle b^\dag\rangle\neq 0$, which further implies that 
$\langle b^\dag b^\dag\rangle\approx \langle b^\dag\rangle^2 \neq 0$. Hence $\Delta$ 
will always be non-zero in the single particle BEC phase. The correct measure of the 
existence of the pairing (independent of the contribution from the single particle BEC) is $\Delta - n_c$. 
We know from the calculation [see Fig.~\ref{fig:OP_alpha}] that for the non-interacting 
Bose gas ($\alpha=0$), $\Delta=n_c$, as it should be. However, for any finite $\alpha$ we find $\Delta-n_c>0$. Thus, for arbitrarily 
small values of the correlated-hopping, the system develops a tendency towards pair formation. However, it does not suppress the single particle BEC in favor of a purely pairing phase until 
sufficiently strong $\alpha$ is reached for sufficiently small $n$. 

The results of our mean-field calculation are similar to those obtained on a different problem in the context of the atomic gases~\cite{Sachdev}. These are studies regarding the 
transition from a purely molecular condensate (MC) to an atomic condensate with a non-zero fraction of the molecular condensate present (AC+MC) accross the Feshbach resonance. 
Our pairing phase is like the MC, and the single particle BEC phase is analogous to the AC+MC. 

The nature of the transition based on symmetry considerations is also similar. The mean-field Hamiltonian, $H_{MF}$, explicitly breaks the U(1) gauge symmetry, 
however it is still invariant under global $Z_2$ symmetry, that is under $b_\r\rightarrow -b_\r$. 
In other words gauge symmetry, $b^\dag_\r\rightarrow b^\dag_\r e^{i\phi}$ leaves $H_{MF}$ 
invariant for $\phi=\pi$. This residual Ising like symmetry will also be broken if there is single 
particle BEC (because $\langle b\rangle\neq 0$). What we have in 
Fig.~\ref{fig:phasediagram} is such an Ising symmetry breaking 
quantum phase transition, where $n_c$ is the relevant order 
parameter. The pairing phase respects this  $Z_2$ symmetry while 
the single particle BEC phase breaks it spontaneously. 

\begin{figure}
\includegraphics[width=7.5cm]{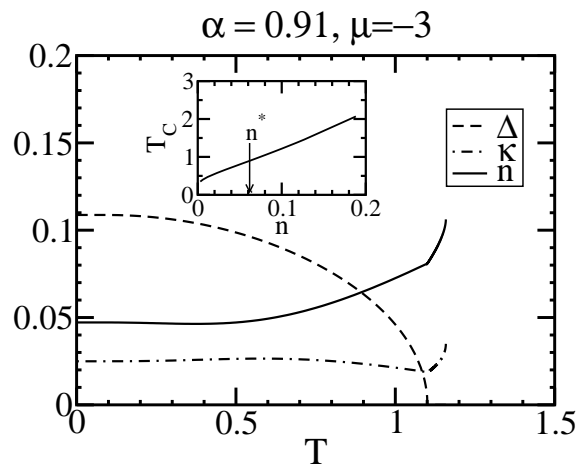}
\caption{The temperature dependence of $\Delta$, $\kappa$ and $n$ for a
 given $\mu$ and $\alpha$. Inset: dependence of $T_c$ on the density.}
\label{fig:OP_T}
\end{figure}

The temperature dependence of various quantities in the mean-field theory is 
shown in Fig.~\ref{fig:OP_T}. The temperature at which $\Delta$ becomes zero 
is called $T_c$. This quantifies the mean-field phase transition from a normal Bose gas at high temperatures to a 
pairing phase below $T_c$. The inset of Fig.~\ref{fig:OP_T} shows $T_c$ as a function of $n$. 
Remarkably, there is no detectable anomaly upon going through the critical
density $n^*$. Since we are in 2D, we do not expect to have a true BEC of pairs, but rather
a Kosterlitz-Thouless (KT) transition. These results suggest that the system should
undergo one KT transition whatever the density, followed by an Ising transition
if $n>n^*$.

To check the validity of the mean-field approximation, hence of our conclusions, it
would be very useful to have unbiased numerical results on the model of Eq. (1).
However, we have good reasons to believe that the predictions of the present mean-field
theory are physically relevant. Mathematically, the structure of the mean-field 
equations and the results of 
the calculations are similar to the Schwinger boson mean-field theory of the quantum 
spin system~\cite{HRK,Mila}. In that context, the single particle BEC phase implies 
an ordered phase in the spin  variables, while the pairing phase denotes a {\em disordered} phase.
Now the physical relevance of these disordered phases is well established in the
context of quantum magnets~\cite{CHN,Sachdev_book}, and we expect the same to be true here.

The model studied in the present paper has similarities with the ring exchange model of bosonic Cooper pairs introduced some 
time ago by Paramekanti and collaborators~\cite{Paramekanti}. In their model, the pairs of bosons hop on the opposite corners of a plaquette. 
The low temperature physics is significantly different however. In their model, the number of bosons is conserved on each row and column 
of the square lattice, leading to the Luttinger liquid like physics and critical correlations in the ground state. In our model, 
the correlated-hopping does not sustain any such conservation law, and the ground state is expected to develop a true long range order.

Finally, let us briefly discuss the physical implications of these results for 
the magnetization process of gapped quantum magnets. The thermodynamics
was already discussed in the boson language: we expect to observe a KT transition
for any magnetization, followed by an Ising transition if the magnetization is larger
than a critical value. This will remain essentially true for 3D systems, the KT
transition being replaced by a true phase transition toward an ordered phase.
However, we also expect very significant differences between the
zero temperature phases. Single particle BEC means magnetic long-range
order, and the system is expected to have gapless transverse spin waves. In other
words, the gap detected in spectroscopies such as inelastic neutron scattering
or NMR will vanish. However, at low magnetization, we only have pair BEC.
The order implied by this pair BEC will be of nematic type since the transverse
components of the spins within a pair can be flipped without changing the 
correlations. But more importantly, there is a gap to single particle excitations,
i.e. to single spin flips. Although the system is gapless in this phase, we thus
expect to observe a gap in neutron scattering or NMR, the gapless excitations
appearing only in the channel $\Delta S = 2$.

In summary, we have shown that the correlated-hopping can change drastically the properties of bosons, leading at low 
densities to a pairing phase without single-particle BEC, and with gapped quasi-particle excitations. In the context of 
quantum frustrated magnets, this leads to the prediction of an Ising phase transition (for low magnetization) as a 
function of the magnetic field, for systems where frustration reduces direct hopping of triplets, thus making correlated-hopping  
the dominant process of kinetic energy.

Beyond frustrated magnets, these results will have implications on all systems where correlated-hopping may be the dominant 
source of kinetic energy. One such class of systems are the atomic gases, where different external parameters control the 
hopping and the Coulomb terms. Whether instabilities of the kind described here can be induced in these systems by reducing 
the single-particle kinetic energy is left for the future investigation.
\begin{acknowledgments}
We thank A. Georges, S. Miyahara, D. Poilblanc and M. Troyer for stimulating discussions on different aspects of this work. 
We also acknowledge the Swiss National Funds and the MaNEP for the financial support.
\end{acknowledgments}
\bibliography{paper}
\end{document}